\documentclass[
	aps,
	twocolumn,
	altaffilletter,
	nolongbibliography,
	numerical,
	pra,
	superscriptaddress,
	floatfix,
	10pt
]{revtex4-1}

\usepackage{graphicx}
\usepackage{amsmath, amssymb}
\usepackage{braket}
\usepackage{newtxtext, newtxmath}
\usepackage{siunitx}

\usepackage[breaklinks, pdftex, hyperfootnotes=true, pdfpagelabels, bookmarks, pageanchor]{hyperref}
\hypersetup{%
	colorlinks=true, linktocpage=true, pdfstartpage=1, pdfstartview=FitH, pdfborder={0 0 0},%
	breaklinks=true, pdfpagemode=UseNone, pageanchor=true, pdfpagemode=UseOutlines,%
	plainpages=false, bookmarksnumbered, bookmarksopen=true, bookmarksopenlevel=1,%
	hypertexnames=true, pdfhighlight=/O,
	urlcolor=blue, linkcolor=blue, citecolor=blue, 
}

\usepackage{lipsum}

\DeclareMathOperator{\Tr}{Tr}
\DeclareMathOperator{\diag}{diag}
\newcommand{\comm}[2]{\left[#1, #2\right]}
\newcommand{\acomm}[2]{\lbrace#1, #2\rbrace}
\newcommand{\gaugesupop}   {\ensuremath{\mathbb{A}}}

\makeatletter
\newsavebox{\@brx}
\newcommand{\llangle}[1][]{\savebox{\@brx}{\(\m@th{#1\langle}\)}%
	\mathopen{\copy\@brx\mkern2mu\kern-0.9\wd\@brx\usebox{\@brx}}}
\newcommand{\rrangle}[1][]{\savebox{\@brx}{\(\m@th{#1\rangle}\)}%
	\mathclose{\copy\@brx\mkern2mu\kern-0.9\wd\@brx\usebox{\@brx}}}
\makeatother

\newcommand*{\kket}        [1]{\ensuremath{\lvert#1\rrangle}}
\newcommand*{\bbra}        [1]{\ensuremath{\llangle#1\rvert}}
\newcommand*{\bbrakket}    [2]{\ensuremath{\llangle#1\vert#2\rrangle}}
\newcommand*{\mmel}        [3]{\ensuremath{\llangle#1\vert#2\vert#3\rrangle}}

\usepackage{comment}

\begin{document}
	
\title{Optimal quantum annealing: A variational shortcut to adiabaticity approach}
	
	\author{G.\,Passarelli}
    \affiliation{CNR-SPIN, c/o Complesso di Monte S. Angelo, via Cinthia - 80126 - Napoli, Italy}
	\author{R.\,Fazio}
	\affiliation{The Abdus Salam International Center for Theoretical Physics (ICTP),, Strada Costiera 11, I-34151 Trieste, Italy}
	\affiliation{Dipartimento di Fisica ``E.\,Pancini'', Universit\`a di Napoli Federico II, Complesso di Monte S.~Angelo, via Cinthia - 80126 - Napoli, Italy}
	\affiliation{Istituto Nanoscienze-CNR, I-56126 Pisa, Italy}
	\author{P.\,Lucignano}
	\affiliation{Dipartimento di Fisica ``E.\,Pancini'', Universit\`a di Napoli Federico II, Complesso di Monte S.~Angelo, via Cinthia - 80126 - Napoli, Italy}

	\begin{abstract}
		Suppressing unwanted transitions out of the instantaneous ground state is a major challenge in unitary adiabatic quantum computation. A recent approach consists in building counterdiabatic potentials approximated using variational strategies. In this contribution, we extend this variational approach to Lindbladian dynamics, having as a goal the suppression of diabatic transitions between pairs of Jordan blocks in quantum annealing. We show that, surprisingly, unitary counterdiabatic ans\"atze are successful for dissipative dynamics as well, allowing for easier experimental implementations compared to Lindbladian ans\"atze involving dissipation. Our approach not only guarantees improvements of open-system adiabaticity but  also enhances the success probability of quantum annealing.
	\end{abstract}
	
\maketitle	
	
\section{Introduction}
\label{sec:intro}

Quantum control of Noisy-Intermediate Scale Quantum (NISQ) processors is an invaluable tool for the design of the ideal time-dependence required to realize accurate  quantum state transformations 
~\cite{sugny:quantum-optimal-control-intro,Werschnik_2007,fazio:quantum-driving,PhysRevA.65.042308,caneva:optimal-control,glaser:schrodinger-cat,koch:2019}.  In quantum many-body systems, new methods of control 
have been developed~\cite{muga2010,doria:2011,choi2015}, with  applications ranging from 
state preparation~\cite{lukin2019} to optimization of hybrid quantum-classical algorithms~\cite{farhi2014,venturelli-qaoa,magann:oct-vqa,an:qoct-reinforcement-learning,wauters:rl-qaoa,mbeng:optimal-digitized-qa}.

Among the different approaches to quantum control, Shortcuts To Adiabaticity (STA) have gained increasing attention~\cite{delcampo:sta-cd,delcampo:assisted-adiabatic-passage,torrontegui:sta,guery-odelyn:sta,campbell:sta,delcampo:quantum-engines,delcampo:qsl-oqs}.
Given a time-dependent Hamiltonian $H_0$, the goal of  STA is to design the evolution of the state  in order to keep the system 
in one of the instantaneous eigenstates of $H_0$. 
A quantum system prepared in an eigenstate of its Hamiltonian at a time $ t = 0 $ will remain in the corresponding time-evolved eigenstate for 
all times even if the conditions for adiabaticity are violated~\cite{guery-odelyn:sta,berry:cd}.  The idea underlying STA is to compensate 
diabatic transitions between instantaneous energy eigenstates by suppressing non-diagonal terms of the Hamiltonian in the energy eigenbasis using 
a counterdiabatic potential. 

While in few-body systems the success of STA has been unquestionable, its application to many-body dynamics has been initially hindered by the fact that exact counterdiabatic operators~\cite{berry:cd,demirplack:2008} are exceedingly 
difficult to realize experimentally, as they often involve non-local infinite-range interactions. In addition, they can only be computed if the full Hamiltonian spectrum is known, a requirement clearly impossible to satisfy in the many-body case. This issue has been solved in a breakthrough 
contribution~\cite{polkovnikov:pnas} by using a variational approach to build approximate counterdiabatic operators. 
In~\cite{polkovnikov:pnas,polkovnikov:nested-commutators,
passarelli:cd-pspin,hartmann:lhz}, this approximate method has been applied successfully to many-body systems.

Applications to real-life problems require generalizing and implementing the variational 
approach to open quantum  systems. This  would be of great importance in the optimization of NISQ protocols in many-body quantum systems. 
A key example of this sort is  quantum annealing (QA)~\cite{albash:rmp,albash:relaxation-qa,chen:pausing-beneficial,kadowaki:thermodynamics-qa,mishra:finite-temperature-qa,
passarelli:reverse-ira,passarelli:dissipative-p-spin,passarelli:pausing,marshall:deciphering-experimental-qa,marshall:pausing,marshall:pausing-2}, where 
the aim is to keep the  system close to its ground state for the entire dynamics up to the annealing time $ \tau $. This procedure will 
eventually lead to the solution of an NP-hard problem~\cite{lucas:nphard,santoro:science}. 
When dissipation is present, the adiabatic time scale $ \hslash / \Delta $, where $ \Delta $ is the 
minimum spectral gap~\cite{Born:1928}, has to be compared with the typical relaxation and decoherence time scales~\cite{albash:decoherence}. As a result, 
the adiabatic theorem  is not sufficient to predict the success probability of QA. The challenge here  
is to find a strategy to 
minimize the effect of environment {\it and},  at the same time, avoid transitions to excited states. 

In many  relevant situations the dynamical evolution is governed by a Lindblad master equation. Quantum annealing is then realized by interpolating a starting and a target Lindbladian, whose zero-temperature instantaneous steady state (ISS) encodes the solution to the problem at hand. Diabatic and thermal transitions outside of the ISS manifold have to be minimized so as to realize high-fidelity quantum computation. Many attempts have been proposed in recent years~\cite{campos-venuti:oc-closed-open,venuti:adiabaticity-oqs,wu:decoherence-free-subspaces,dupays:superadiabatic-thermalization,dann:ste-oqs,wu:sta-for-oqs} for related questions in unitary evolution~\cite{susa:exponential-speedup,nishimori:non-stoq,yamashiro:ara,passarelli:reverse-ira}.

An important leap forward in the field entails exploiting STA to advantageously design nonadiabatic protocols in the presence of dissipation. 
STA in open quantum systems have been studied in Refs.~\cite{vacanti:transitionless-open-systems,Alipour:2020shortcutsto}. 
In this work, we focus on the  quantum many-body case generalizing the variational approach of Ref.~\cite{polkovnikov:pnas} to open systems.
In particular, we derive a variational approach for Lindbladian dynamics and apply it to 
quantum annealing. This allows for the optimization of quantum driving in real-life scenarios by building approximate Lindbladian CD operators which satisfy 
locality constraints, in order to match with the actual experimental capabilities. 
Remarkably,  in many relevant cases, it is sufficient to control  the unitary part of the counterdiabatic terms. These are much 
more easily engineered experimentally and already realize good approximations of the exact counterdiabatic superoperator. We show that 
our control on the Hamiltonian considerably increases the ground state fidelity, therefore boosting the performance of quantum annealing.
	
This paper is organized as follows. In Sec.~\ref{sec:transitionless-lindblad}, we discuss exact CD driving in Lindbladian dynamics adopting a superoperator 
representation~\cite{Alicki:book}. In Sec.~\ref{sec:variational}, we present our variational approach of the search for the open-system CD superoperator. Our approach resembles the unitary case of Ref.~\cite{polkovnikov:pnas}, however there are important differences 
due  to the fact that the generators of the dynamics are not Hermitian. 
In Sec.~\ref{sec:results}, firstly we validate our analysis by applying our formalism to a single qubit in interaction with an Ohmic environment, and, secondly, we study the ferromagnetic $ p $-spin model with $ p = 3 $ 
as a paradigmatic example of a quantum annealing protocol of a many-body system. We draw our conclusions in Sec.~\ref{sec:conclusions}.

\section{Transitionless Lindbladian dynamics} 
\label{sec:transitionless-lindblad}

Let us consider a quantum state $ \ket{\psi(t)} $ evolving according to the Schr\"odinger equation with a time-dependent Hamiltonian $ H_0(t) $. 
At any time, the state $ \ket{\psi(t)} $ can be decomposed as $ \ket{\psi(t)} = \sum_\alpha c_\alpha(t) \ket{d_\alpha(t)} $ where  $ \ket{d_\alpha(t)} $ are instantaneous eigenvectors of $H_0(t)$. 
The adiabatic regime is realized for long time scales $\tau\gg\max\{\hbar/[\epsilon_\alpha(t)-\epsilon_\beta (t)]\}$, where $\epsilon_\alpha(t)$ are the instantaneous eigenvalues of $H_0(t)$.  
In this limit, each coefficient $ c_\alpha(t) $ evolves independently of the others. As shown in Refs.~\cite{berry:cd,demirplack:2008} the same results can be achieved for any finite time $\tau$, provided the generator of the dynamics is $ H(t) = H_0(t) + H_\text{cd}(t) $, where
\begin{equation}\label{eq:cd-unitary}
    H_\text{cd}(t) = i\hslash \sum_{\alpha\ne\beta} \frac{\langle{d_\beta(t)}\vert{\dot{H}_0(t)}\vert{d_\alpha(t)}\rangle}{\epsilon_\alpha(t) -\epsilon_\beta(t)} \lvert{d_\beta(t)}\rangle\langle{d_\alpha(t)}\rvert,
\end{equation}
is the so called CD potential.

In all practical scenarios where dissipation is present, the evolution of a quantum system is not unitary but can instead be modeled by a Lindblad equation for the reduced density operator $\rho(t)$
\begin{equation} 
	\dot\rho(t) = \mathcal{L}(t)[\rho(t)] \;,
\end{equation}
where $ \mathcal{L}(t)[\bullet] $ is a (time-dependent) completely positive trace preserving (CPTP) map expressed in the Lindblad form
\begin{equation}
	\mathcal{L}(t)[\rho] = -i \comm{H(t)}{\rho} + \sum_k \gamma_k(t) \left(L_k \rho L_k^\dagger - \frac{1}{2}\acomm{L_k^\dagger L_k}{\rho}\right)
\end{equation}
where $ H(t) $ is the Hermitian generator of the unitary part of the evolution and $ L_k $ are Lindblad operators with rates $ \gamma_k(t) \ge 0 $ for all $t$.

In the superoperator representation \cite{sarandy:approximation}, first a basis $ \set{\sigma_i}_{i = 0}^{D^2-1} $ of Hermitian and traceless operators (including the identity $i=0$) is chosen, where $ D $ is the Hilbert space dimension. This defines the Hilbert-Schmidt space, whose vectors $ \sigma_i \leftrightarrow \kket{\sigma_i} $ are orthonormal with respect to the scalar product $ (\sigma_i, \sigma_j) \equiv \bbrakket{\sigma_i}{\sigma_j} = \Tr(\sigma_i \sigma_j) / D $. In this basis, density matrices are represented as $ D^2 $-dimensional coherent vectors $ \kket{\rho} = \sum_i r_i \kket{\sigma_i} $ with coefficients  $ r_i = \bbrakket{\sigma_i}{\rho} $ and the Lindblad equation takes the form 
\begin{equation}
\label{lind1}
	\kket{\dot{\rho}} = \mathbb{L}_0(t) \kket{\rho},
\end{equation} 
where $ \mathbb{L}_0(t) $ is the Lindbladian supermatrix having components $ [\mathbb{L}_0]_{jk} = \bbrakket{\sigma_j}{\mathcal{L}(t)[\sigma_k]} $. In this paper, bold symbols are used to indicate supermatrices and double kets (or bras) are reserved for coherence vectors, whereas calligraphic letters indicate the action of superoperators in the original Hilbert space.
	
While formally identical to the Liouville equation, Eq.~\eqref{lind1}  generates a nonunitary evolution since, in general, $ \mathbb{L}_0(t) $ is not anti-Hermitian and non-diagonalizable. However, it can  always be brought to the Jordan canonical form (JF), which is unique up to permutations~\cite{nering:1970}. In this form, $ \mathbb{L}_0(t) $ assumes a block-diagonal structure $ \mathbb{L}_\text{J}(t) = \mathbb{V}^{-1}(t) \mathbb{L}_0(t) \mathbb{V}(t) = \diag[J_0(t), \cdots, J_{N-1}(t)]  $, where $ \mathbb{V}(t) $ is a similarity matrix and each Jordan block (JB) $ J_\alpha(t) $ is given by $ {[J_\alpha(t)]}_{ij} = \lambda_\alpha(t) \delta_{ij} + \delta_{i,j+1} $.
Each JB is associated with different non-crossing time-dependent (complex) eigenvalues of $ \mathbb{L}_0(t) $, denoted $ \lambda_\alpha(t) $. Each eigenvalue has algebraic multiplicity $ N_\alpha $ and geometric multiplicity equal to one.  If the Lindbladian has exactly $ D^2 $ eigenvectors, each JB is a one-dimensional matrix and $ \mathbb{L}_\text{J}(t) $ is diagonal. This is the 1D Jordan form (1DJF). In this case, the Lindblad supermatrix is diagonalizable with complex eigenvalues.
	
Given the above, $ \mathbb{L}_0(t) $ does not generally yield a basis of eigenvectors. However, we can define a basis of right and left quasi-eigenvectors that solve the following problems:  
\begin{align} 
	\mathbb{L}_0(t) \kket{\mathcal{D}_\alpha^{n_\alpha}} &= \kket{\mathcal{D}_\alpha^{n_\alpha-1}} + \lambda_\alpha(t) \kket{\mathcal{D}_\alpha^{n_\alpha}},\\
	\bbra{\mathcal{E}_\alpha^{n_\alpha}}\mathbb{L}_0(t) &= \bbra{\mathcal{E}_\alpha^{n_\alpha+1}} + \lambda_\alpha(t) \bbra{\mathcal{E}_\alpha^{n_\alpha}} 
\end{align} 
with $ n_\alpha \in\{ 1, \dots, N_\alpha\} $ ($ \kket{\mathcal{D}_\alpha^{0}} $ and $ \bbra{\mathcal{E}_\alpha^{N_\alpha+1}} $ are null vectors). $ N_\alpha $ 
is the algebraic multiplicity of the eigenvalue $ \lambda_\alpha $. These states are doubly orthogonal ($ \bbrakket{\mathcal{E}_\beta^{n}}{\mathcal{D}_\alpha^{m}} = \delta_{mn} \delta_{\alpha\beta} $) and decompose the identity as $ \mathbb{I} = \sum_{\alpha = 0}^{N - 1} \sum_{n_\alpha=1}^{N_\alpha} \kket{\mathcal{D}_\alpha^{n_\alpha}}\bbra{\mathcal{E}_\alpha^{n_\alpha}} $. In the 1DJF, these quasi-eigenstates become exact eigenstates ($ N_\alpha = 1 $).

The dynamics 
of an open quantum system is said to be adiabatic when the evolution of the density operator in its Hilbert-Schmidt space 
can be decomposed into decoupled Jordan subspaces associated with distinct, time-dependent, non-crossing eigenvalues of the Lindbladian 
supermatrix $ \mathbb{L}_0(t) $~\cite{sarandy:approximation,santos:adiabaticity-open}. 
For any finite evolution time $\tau$, it is possible to define transitionless dynamics generated by $ \mathbb{L}(t) = \mathbb{L}_0(t) + \mathbb{L}_\text{cd}(t) $ 
introducing a counterdiabatic superoperator $ \mathbb{L}_\text{cd}(t) $ that ensures Jordan blocks are not mixed. The detailed derivation of 
$ \mathbb{L}_\text{cd}(t) $ can be found in Refs.~\cite{sarandy:approximation,vacanti:transitionless-open-systems}. For 1DJF, it reads
	\begin{equation}
	\label{eq:cd-open}
	\mathbb{L}_\text{cd}(t) = \sum_{\alpha\ne\beta}  \frac{\bbra{\mathcal{E}_\beta(t)}\dot{\mathbb{L}}_{0}(t)\kket{\mathcal{D}_\alpha(t)}}{\lambda_\alpha(t)-\lambda_\beta(t)}\kket{\mathcal{D}_\beta(t)}\bbra{\mathcal{E}_\alpha(t)},
	\end{equation}
which generalizes the unitary CD operator~\cite{demirplack:2008,berry:cd} to the 1DJF open case with $ N_\alpha = 1 $, where $ \bbra{\mathcal{E}_\alpha(t)} $ and 
$ \kket{\mathcal{D}_\alpha(t)} $ are the left and right eigenvectors of $ \mathbb{L}_0(t) $ with eigenvalue $  \lambda_\alpha(t) $, 
respectively~\cite{vacanti:transitionless-open-systems}.  This formula is derived under the assumption that the Lindbladian spectrum is non-degenerate so that the denominators are non-zero. 
The explicit form of the CD Lindbladian in the case the JF is not 1D can be found in Ref.~\cite{sarandy:approximation}.
    
Evaluating the open-system CD superoperator would require the knowledge of the 
whole Lindbladian spectrum, a requirement that it is impossible to fulfill in many-body systems.

\section{Variational formulation}
\label{sec:variational}

First of all, we quickly review the variational formulation of the search for approximate CD operators in closed quantum systems. If the system Hamiltonian depends on time via a parameter $ s = s(t) $, then the CD potential of Eq.~\eqref{eq:cd-unitary} can be equivalently~\cite{polkovnikov:pnas} written as $ H_\text{cd}(t) = \dot{s}(t) A_s $, and the gauge potential $ A_s $ satisfies 
\begin{equation}
\label{eq:gauge-potential-unitary}
	[i\hslash H_0'(s) - [A_s,H_0(s)],H_0(s)] = 0,
\end{equation}
where the prime denotes derivative with respect to $ s $. Solving this equation is equivalent to minimizing the functional $ S(A_s^*) = \Tr[G_s^2(A_s^*)] $ with respect to $ A_s^* $, where $ G_s(A_s^*) = H_0'(s) + i [A_s^*,H_0(s)] / \hslash $. 
At this point one could express $ A^*_s $ as a combination of local operators and  minimize the functional $ S(A^*_s) $ to achieve an approximation of the CD potential.  This variational approach to the CD driving, in the unitary case, has been successfully applied, achieving high-fidelity transitionless quantum driving, in Refs.~\cite{polkovnikov:pnas,polkovnikov:nested-commutators,passarelli:cd-pspin,hartmann:lhz}.

We here present a variational approach to approximate $\mathbb{L}_\text{cd}$  without 
knowing the Lindbladian spectrum, generalizing Ref.~\cite{polkovnikov:pnas} to open quantum systems. Assuming the Lindbladian depends on time via the control field $ s = s(t) $, we can 
rewrite Eq.~\eqref{eq:cd-open} as $ \mathbb{L}_\text{cd}(t)= \dot{s}\gaugesupop_s $, where $ \gaugesupop_s $ is the counterdiabatic 
gauge supermatrix. To formulate the search for $ \mathbb{A}_s $ on a variational basis, we start from the Lindblad equation and rotate to the Jordan 
representation using the similarity superoperator $ \mathbb{V}(t) $. The system density matrix in this basis reads $ \kket{\tilde{\rho}(t)} = 
\mathbb{V}^{-1}(t) \kket{\rho(t)} $ and satisfies the adiabatic-frame Lindblad equation 
	\begin{equation}
	\partial_t \kket{\tilde{\rho}(t)} = \left(\mathbb{L}_\text{J} - \tilde{\mathbb{L}}_\text{cd} - \tilde{\mathbb{L}}_\text{d}\right) \kket{\tilde{\rho}(t)},
	\end{equation}
where $ \mathbb{L}_\text{J}(t) = \mathbb{V}^{-1} \mathbb{L}_0(t) \mathbb{V} = \diag[J_0,(t)\cdots, J_{N-1}(t)] $ and $ J_\alpha $ is the Jordan block 
associated with the eigenvalue $ \lambda_\alpha $~\cite{Note1}, and $ \tilde{\mathbb{L}}_\text{cd}(t) + 
\tilde{\mathbb{L}}_\text{d}(t)  = -\partial_t (\mathbb{V}^{-1}) \mathbb{V} = \mathbb{V}^{-1} \partial_t\mathbb{V} $.	We have defined
	\begin{align}
	\tilde{\mathbb{L}}_\text{cd}(t) &= \sum_{\alpha\ne\beta} \sum_{n_\alpha n_\beta} \kket{\mathcal{D}_{\alpha}^{n_\alpha}} \bbrakket{\mathcal{E}_{\alpha}^{n_\alpha}}{\partial_t \mathcal{D}_{\beta}^{n_\beta}}\bbra{\mathcal{E}_{\beta}^{n_\beta}} \\ 
	\tilde{\mathbb{L}}_\text{d}(t) &= \sum_{\alpha} \sum_{n_\alpha n_\beta} \kket{\mathcal{D}_{\alpha}^{n_\alpha}} \bbrakket{\mathcal{E}_{\alpha}^{n_\alpha}}{\partial_t \mathcal{D}_{\alpha}^{n_\beta}}\bbra{\mathcal{E}_{\alpha}^{n_\beta}}. 
	\end{align} 
Note that the  superoperator $ \tilde{\mathbb{L}}_\text{d}(t) $ is only diagonal in the Jordan indices but not necessarily within each Jordan block.

It is easy to prove that the CD superoperator satisfies 
    \begin{equation}
         \partial_t \mathbb{L}_0 + \mathbb{F}_\text{ad}(t) = \comm{\mathbb{L}_\text{cd}(t) + \mathbb{L}_\text{d}(t)}{\mathbb{L}_0(t)},
    \end{equation} 
where quantities without the tilde are in the time-dependent basis and 
    \begin{equation}
    \mathbb{F}_\text{ad}(t) = -\mathbb{V} \partial_t \mathbb{L}_\text{J}(t) \mathbb{V}^{-1} = -\sum_{\alpha n_\alpha} \partial_t\lambda_\alpha \kket{\mathcal{D}_{\alpha}^{n_\alpha}} \bbra{\mathcal{E}_{\alpha}^{n_\alpha}}. \end{equation}
    %

This equation is equivalent to
    \begin{equation}\label{eq:gauge-potential-open-general}
    	\comm{\mathbb{L}_0'(s) - \comm{\mathbb{A}_s + \mathbb{A}_\text{d}}{\mathbb{L}_0(s)}}{\mathbb{L}_0(s)} = 0,
    \end{equation}
	where $ \mathbb{A}_\text{d} $ is defined via the equation $ \mathbb{L}_\text{d} = \dot s \mathbb{A}_\text{d} $. Eq.~\eqref{eq:gauge-potential-open-general} is formally equivalent to the unitary case of Ref.~\cite{polkovnikov:pnas}. 
 However, the fact that the Jordan basis is not unitarily equivalent to the time-independent basis $ \set{\sigma_i} $ poses a problem. Inspired by the unitary case, 
 we can define a superoperator $ \mathbb{G}_s(\mathbb{A}_\text{tot}^*) = \mathbb{L}_0'(s) - [\mathbb{A}_\text{tot}^*,\mathbb{L}_0(s)] $ with $ \mathbb{A}_\text{tot}^* = \mathbb{A}_s^* + \mathbb{A}_\text{d}^* $ and notice that 
 its diagonal elements in the Jordan basis do not depend on $ \mathbb{A}_\text{tot}^* $ while its off-diagonal elements are zero when 
 $ \mathbb{A}_\text{tot}^* = \mathbb{A}_\text{tot} = \mathbb{A}_s + \mathbb{A}_\text{d} $. Thus, the Hilbert-Schmidt norm of $ \mathbb{G}_s $ in the Lindbladian basis is minimized when 
 $ \mathbb{A}_\text{tot}^* = \mathbb{A}_\text{tot} $. However, this approach does not provide any insight as the trace in $ \mathbb{S} = \Tr(\mathbb{G}_s^\dagger \mathbb{G}_s) $ 
 has to be evaluated in the Jordan eigenbasis, requiring the full diagonalization of the Lindbladian.
    
Hence, we work directly with Eq.~\eqref{eq:gauge-potential-open-general}. An approximate solution to Eq.~\eqref{eq:gauge-potential-open-general} might lead to a total generator of the system dynamics $ \mathbb{L}(s) = \mathbb{L}_0(s) + \dot{s} \mathbb{A}_s $ that does not yield completely-positive trace-preserving (CPTP) dynamics. This problem can be circumvented by simply restricting the variational minimization to a subspace of the supermatrix space that contains only physically valid superoperators. In particular, the Kraus representation theorem~\cite{breuer:open-quantum} and the works by \textcite{gorini:completely-positive-dynamics} and \textcite{lindblad:complete-positivity} ensure that for finite-size or separable Hilbert spaces the most general form of a CPTP map is given by
	\begin{align}\label{eq:gauge-potential-lindbladian}
		&\mathcal{A}_\text{tot}^*[\bullet] = -\frac{i}{\hslash} \comm{A_s^*}{\bullet} \notag\\&\quad+ \sum_i \gamma_i^2 \left(\Gamma_i \bullet \Gamma_i^\dagger - \frac{1}{2}\acomm{\Gamma_i^\dagger \Gamma_i}{\bullet}\right), 
	\end{align}
where $ {(A^*)}^\dagger = A_s $ and $ \Gamma_i $ are Lindblad operators. By postulating this form, we can expand the superoperator $ \gaugesupop_\text{tot}^* $ as
\begin{equation}
    \gaugesupop_\text{tot}^* = \sum_i \epsilon_i \gaugesupop_i^\text{uni} + \sum_i \gamma_i^2 \gaugesupop^\text{diss}_i \equiv \sum_j \alpha_j \gaugesupop_j,
\end{equation}
where we separated the unitary and dissipative contributions to $\gaugesupop_\text{tot}^*$ so as to impose the positivity of the rates in the Lindblad form of Eq.~\eqref{eq:gauge-potential-lindbladian}. The map of Eq.~\eqref{eq:gauge-potential-lindbladian} would require bath engineering, which might be daunting in practice. However, as proven in Ref.~\cite{vacanti:transitionless-open-systems}, there exist cases (for example when the system purity is unaffected by the dynamics) where the unitary part of this map is sufficient to express the exact CD superoperator of the Lindbladian dynamics, providing an important simplification. Furthermore, in Appendix~\ref{app:cd-uni-from-open}, we show that the closed-system variational formulation can be obtained as a limiting case of our, more general approach when all maps are unitary.

%
%
Eq.~\eqref{eq:gauge-potential-open-general} can then be recast into 
$ \vec{\mathbb{B}} \cdot \vec{X} = \mathbb{Y} $, where $ X_i = \alpha_i $, $ \mathbb{B}_{i} = \comm{\comm{\mathbb{A}_i}{\mathbb{L}_0}}{\mathbb{L}_0} $, 
and $ \mathbb{Y} = \comm{\mathbb{L}_0'}{\mathbb{L}_0} $. Therefore, the search for $ \alpha_i $ can be reformulated as a minimization problem of the form
    \begin{equation}\label{eq:variational-open-naive-minimization}
    	\vec{X} = \arg\min_{\vec{X}^*} {\lVert\,\vec{\mathbb{B}} \cdot \vec{X}^* - \mathbb{Y}\,\rVert}^2,
    \end{equation}
where the norm is defined as $ \lVert \mathbb{Q} \rVert \equiv \Tr({\mathbb{Q}^\dagger \mathbb{Q}}) $. The outcome of this minimization does not depend on the choice of the starting state.

	\begin{figure*}[t]
		\centering
		\includegraphics[width = 0.9\linewidth]{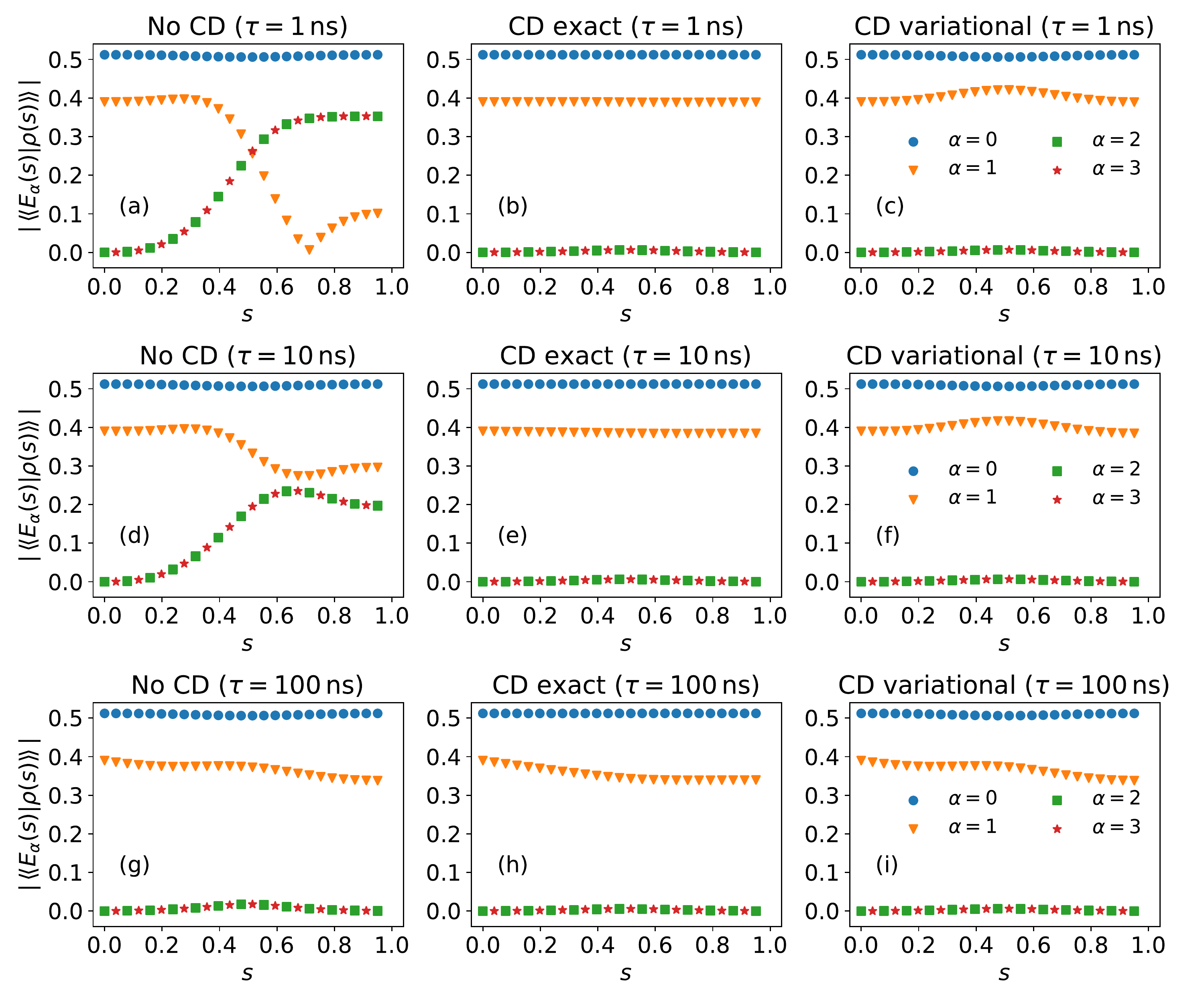}
		\caption{Jordan blocks overlaps $ {\lvert\bbrakket{\mathcal{E}_{\alpha}(s)}{\rho(s)}\rvert} $ as a function of $ s = t / \tau $, for a single qubit in interaction with an Ohmin environment. In panels (a), (d), and (g), we show results obtained without the CD superoperator. In panels (b), (e), and (h), we show results obtained with the exact CD superoperator of Eq.~\eqref{eq:cd-open}. In panels (c), (f), and (i), we show results obtained with the variational CD superoperator $ \mathcal{A}_s[\bullet] = -i \comm{y \sigma_y}{\bullet} $, where $y$ is the variational parameter found by minimizing Eq.~\eqref{eq:variational-open-naive-minimization}. Panels (a--c): $\tau = \SI{1}{\nano\second}$; Panels (d--f): $ \tau = \SI{10}{\nano\second} $; Panels (g--i): $\tau = \SI{100}{\nano\second}$.}
		\label{fig:single-qubit-dissipative-jb-all-tfs}
	\end{figure*}

Lindbladians having a 1DJF are more likely to appear when modeling dissipative systems using weak-coupling master equations, therefore in the following we will focus on this case, where the superoperator $ \mathbb{A}_\text{d} $ commutes with $ \mathbb{L}_0(s) $ and Eq.~\eqref{eq:gauge-potential-open-general} simplifies to 
\begin{equation}\label{eq:gauge-potential-open}
	\comm{\mathbb{L}_0'(s) - \comm{\mathbb{A}_s}{\mathbb{L}_0(s)}}{\mathbb{L}_0(s)} = 0.
\end{equation}

First, we validate our method by applying to a single qubit in interaction with an Ohmic environment. Then, we will focus on the ferromagnetic $p$-spin model, a prototypical many-body system that will showcase the power of our method to improve quantum annealing of complex systems. 

\section{Results}
\label{sec:results}

\subsection{Single qubit in an Ohmic environment} 
\label{subsec:single-qubit}

We consider a single qubit in interaction with a thermal Ohmic bath. The qubit Hamiltonian reads 
\begin{equation}
H_0(s) = -[1-q(s)](\omega_x/2) \sigma_x - q(s) (\omega_z/2) \sigma_z  
\end{equation} 
where $ \omega_x = \omega_z = \SI{1}{\giga\hertz}$
is the energy scale (in units in which $ \hslash = 1 $) and $ q(s) = 6 s^5 - 15 s^4 + 10 s^3 $ with $ s = t/\tau $. The minimum gap is $ \Delta = 1/\sqrt{2} $ at $ s = q = 1/2 $. 

\begin{figure*}[t]
	\centering
	\includegraphics[width = \linewidth]{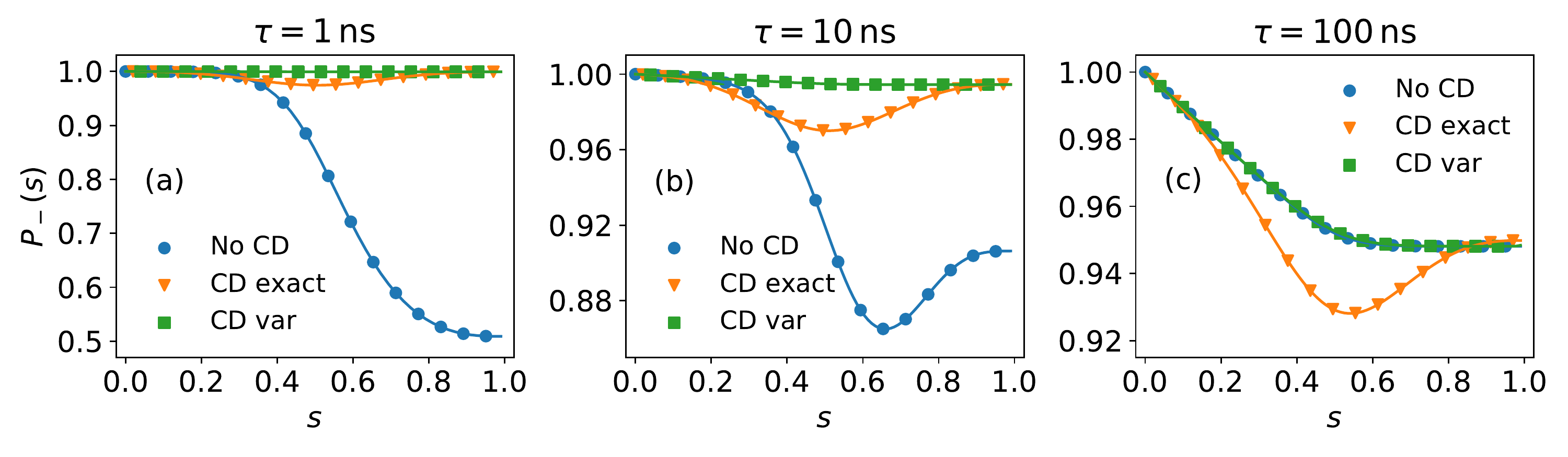}
	\caption{Ground state probability as a function of $ s = t / \tau $, for a single qubit in interaction with an Ohmic environment. Panel (a): $\tau = \SI{1}{\nano\second} $; Panel (b): $\tau = \SI{10}{\nano\second} $; Panel (c): $\tau = \SI{100}{\nano\second} $. Note the different scale on the vertical axis.}
	\label{fig:single-qubit-dissipative-p0-all-tfs}
\end{figure*}	

The  gauge potential in the absence of system bath coupling can be found analytically using Eq.~\eqref{eq:gauge-potential-open} and  reads 
\begin{equation}
\label{eq:gaugep_singleqb}
    A_s = -q'(s)/(1-2q(s)+2q^2(s)) \, \sigma_y.
\end{equation}
Let us now assume that the system is  weakly coupled to a dissipative bath and can be described by a Lindblad master equation~\cite{zanardi:master-equations,albash:decoherence}.  The system is coupled to the bath via $ U = \sigma_z $, which induces dephasing, where the details of the system-bath coupling and of the Lindblad equation are reported in the Appendix~\ref{app:lindblad-equation}. We consider a temperature of $ T = \SI{17}{\milli\kelvin} = \SI{2.23}{\giga\hertz} $ and a dimensionless qubit-bath coupling strength of $ \eta g^2 = \num{1e-4} $. The weak coupling Lindbladian has a zero eigenvalue ($ \lambda_0 = 0 $) for any given $ t $, whose right eigenvector $ \kket{\mathcal{D}_0(t)} $ is the instantaneous steady state (ISS) and corresponds to the thermal state $ \exp(-\beta H(t)) / \Tr [\exp(-\beta H(t))] $. The Lindbladian has a 1DJF for every choice of parameters, thus there are $ D^2 = 4 $ Jordan blocks whose eigenvalues are known explictly~\cite{venuti:adiabaticity-oqs}. 

We initialize the qubit in the ground state of the Hamiltonian at $ s = 0 $, i.\,e., the eigenstate of $ \sigma_x $ with eigenvalue $ +1 $. Then, we fix an annealing time $ \tau $ and evolve the system according to the Lindblad equation. We calculate the ground state probability and also the projections of the time-evolved state onto the four Jordan blocks. When the unitary evolution is adiabatic, the populations of the energy eigenstates stay constant in time. This is a consequence of the fact that each eigenstate will only acquire a phase factor. As opposed to the unitary case, however, open-system adiabaticity does not imply constant projections onto the Jordan blocks, due to the fact that the eigenvalues of the Lindbladian are complex in general. In particular, only the projection onto the ISS will remain constant in time if the adiabatic condition is met, whereas the numerical values of the other blocks might change. However, if some of the populations are zero at the beginning of the dynamics, they are bound to remain zero for the entire adiabatic dynamics and this is a measure of adiabaticity.

We consider three different values of $ \tau $, corresponding to three different regimes: $ \tau = \text{\SIlist{1;10;100}{\nano\second}} $. The value of $ \tau = \SI{1}{\nano\second} $ is the quench limit: in the absence of CD terms unitary and  dissipative dynamics overlap, as the time is too short with respect to the typical time scales of the bath.  For $ \tau = \SI{10}{\nano\second} $, the  fidelity in the unitary limit (calculated as the ground state probability at the time $\tau$) is close to $91\%$, and is similar in the presence of the environmental bath ($90\%$) since the evolution time is again shorter than the relaxation time scale. 
In addition, the dissipative adiabatic criterion is violated, and thus we cannot follow each Jordan block adiabatically in the absence of a CD driving term. Indeed, it has been proven in Ref.~\cite{venuti:adiabaticity-oqs} that, in order to follow the ISS with a maximum error in the norm of the evolved state of $ \epsilon $ when $ U = \sigma_z $, the annealing time must be chosen as $ \tau \gtrsim C / \epsilon^3 $ where $ C $ is a positive constant depending on the specific norm used. For $ \tau = \SI{100}{\nano\second} $, the time evolution is almost adiabatic as shown in the following, and the CD plays a marginal role. Here, the annealing time is larger than the relaxation time scale and the ground state probability at the end of a dissipative evolution drops to $95\%$, as opposed to the unitary limit in which it is close to $100\%$ up to numerical errors.

As a variational ansatz, we take inspiration from the analytic unitary result of Eq.~\eqref{eq:gaugep_singleqb} and consider $ \mathbb{A}_s \to \mathcal{A}_s^\text{test}[\bullet] = -i\comm{y \sigma_y}{\bullet} $, where $ y $ is the variational parameter to be optimized minimizing Eq.~\eqref{eq:variational-open-naive-minimization}.

In Fig.~\ref{fig:single-qubit-dissipative-jb-all-tfs}, we plot the populations of each JB  $ {\lvert\bbrakket{\mathcal{E}_{\alpha}(s)}{\rho(s)}\rvert} $ as a function of the dimensionless time $ s = t / \tau $. In the left-hand column, the qubit has been evolved by using $ \mathbb{L}_0 $ alone without including CD corrections. The center column shows the same quantities when the qubit has been evolved using $ \mathbb{L}_0 + \dot{s} \mathbb{A}_s $ with the exact CD superoperator computed using Eq.~\eqref{eq:cd-open}. The right-hand column shows the populations when the qubit has been evolved using $ \mathbb{L}_0 + \dot{s} \mathbb{A}_s^\text{test} $. When there is no CD term, the populations of the last two Jordan blocks quickly grow as the two subspaces mix with the second block due to nonadiabatic transitions when $ \tau $ violates the dissipative adiabatic condition. By contrast, these populations remain small when $ \tau = \SI{100}{\nano\second} $. The exact CD potential decouples the dynamics of the last two JBs and the populations of these levels remain zero for all choices of $ \tau $. The population of the block with $ \alpha = 1 $ slightly decreases as a consequence of the fact that the (real) eigenvalue $ \lambda_1(t) $ is small and negative. The variational CD superoperator successfully decouples the dynamics of the different blocks as well.
	
Adding the dissipative CD superoperator to the Lindbladian $ \mathbb{L}_0 $ increases the ground state probability as well. 
This is due to the fact that the target state is decomposed onto the first two Jordan blocks at $ s = 1 $ as well, thus the suppression of nonadiabatic transitions outside of these Jordan blocks allows one to reach the target state with more accuracy. This is shown in Fig.~\ref{fig:single-qubit-dissipative-p0-all-tfs}, where we plot the ground state probability $ P_-(s) = \Tr[\ket{d_0(s)}\bra{d_0(s)} \rho(s)] $ as a function of $ s $, where $\ket{d_0(s)} $ is the instantaneous ground state of $ H_0(s) $.
	
\subsection{Ferromagnetic \texorpdfstring{$p$}{p}-spin model}
\label{subsec:p-spin}

Next, we consider the ferromagnetic $ p $-spin model, whose Hamiltonian reads
\begin{equation}\label{eq:pspin}
H_p(s) = -\Gamma [1-q(s)] \sum_{i=1}^{n} \sigma_i^x -\frac{J}{n^{p-1}} q(s) {\left(\sum_{i=1}^{n} \sigma_i^z\right)}^p     
\end{equation}
with $ \Gamma = J = \SI{1}{\giga\hertz} $ and $ q(s) = 6 s^5 - 15 s^4 + 10 s^3 $ with $ s = t/\tau $. We consider $ n = 3 $ qubits with $ p = 3 $. The unitary dynamics of this system is easy to simulate due to the fact 
that the $ p $-spin Hamiltonian commutes with the total angular momentum $ S^2 = S_x^2 + S_y^2 + S_z^2 $ with $ 2S_\alpha = \sum_i 
\sigma_i^\alpha $ ($ \alpha \in \set{x, y, z} $) at all times. In addition, the interesting states for QA, i.\,e., the paramagnetic ground state of 
$ S_x $ and the ferromagnetic ground state of $ S_z^p $, both belong to the symmetry subspace corresponding to $ S = n / 2 $ and 
$ D = 2S + 1 = n + 1 $, therefore numerical simulations of any unitary dynamics can be restricted to this $ D $-dimensional space. 

\begin{figure}[t]
	\centering
	\includegraphics[width = \columnwidth]{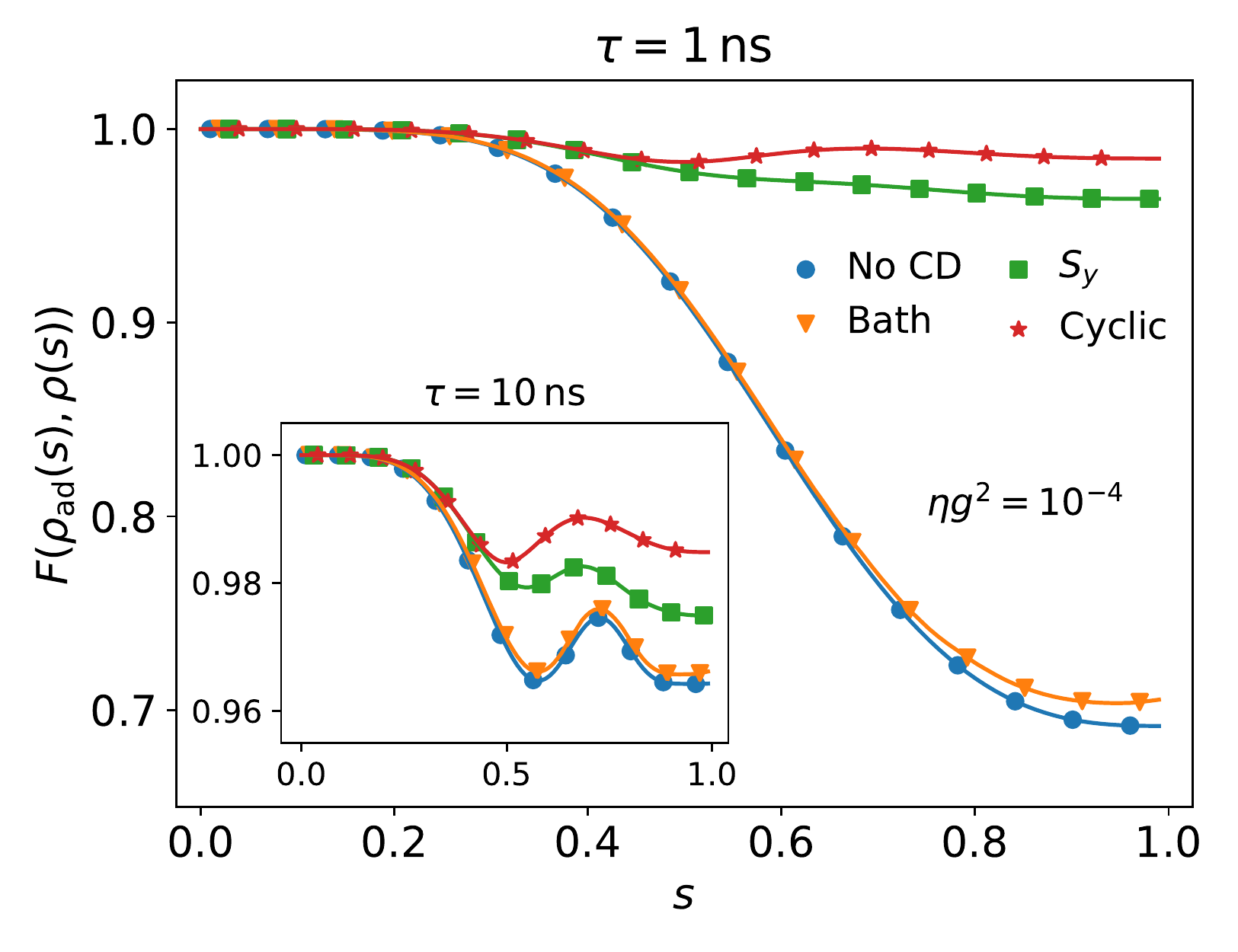}
	\caption{Fidelity between the time-evolved density matrix $ \rho(s) $ and the thermal state $ \rho_\text{ad}(s) $ as a function of $ s = t/\tau $, for $\eta g^2 = \num{1e-4}$. The different curves refer to the three ans\"atze described in the main text.}
	\label{fig:pspin-dissipative-fidelitydm-tf-1-10-ansatz-contributions-eta-0-0001}
\end{figure}

To preserve this symmetry, we consider a collective dephasing model where the whole system is collectively coupled to a single 
dephasing Ohmic bath via the total magnetization $ S_z $. This form can appear experimentally when a qubit system is coupled 
to a long-wavelength mode of the bath, so that the qubit system is insensitive to spatial variations of the bath 
modes~\cite{passarelli:pausing,passarelli:dissipative-p-spin}. The coupling to the environment is modeled via the adiabatic master 
equation~\cite{zanardi:master-equations}. We consider a temperature of $ T = 1/\beta = \SI{17}{\milli\kelvin} = \SI{2.23}{\giga\hertz} $ 
and dimensionless qubit-bath coupling strengths of $ \eta g^2 = \text{\numlist{1e-4;1e-2}} $. 

For $ n = 3 $, the Hilbert space dimension is $ D = 4 $. The $ 4 \times 4 $ operator space is spanned by the basis of operators $ \set{\Sigma_i}_{i=0}^{D^2-1} $, where $ \Sigma_0 $ is the identity and the remaining $ D^2 - 1 $ operators are Hermitian, traceless, and orthonormal. In particular, the basis operators are ($ m > l $) $ {(\Sigma_{2i-1})}_{lm} = \sqrt{2} \delta_{l,m+1} + \sqrt{2} \delta_{l+1,m} $ and $ {(\Sigma_{2i})}_{lm} = -i\sqrt{2} \delta_{l,m+1} + i\sqrt{2} \delta_{l+1,m} $ with $ i = (m-l) + (l-1)(2D-l)/2 \in \set{1,2,\dots,6} $. The remaining three operators are $ \Sigma_{13} = \diag(1,1,-1,-1) $, $ \Sigma_{14} = \diag(1,-1,1,-1) $, and $ \Sigma_{15} = \diag(1,-1,-1,1) $. The Lindbladian is diagonalizable and there are $ D^2 = 16 $ 1D Jordan blocks. 

We prepare the $p$-spin system into the thermal state $ \kket{\mathcal{D}_0(0)} $, i.\,e., the starting density matrix is $ \rho(0) = \exp(2\beta \, \Gamma S_x) / \Tr [\exp(2\beta \,\Gamma S_x)] $. For $ \beta\to\infty $, this state corresponds to the ground state of the Hamiltonian in Eq.~\eqref{eq:pspin}. At the beginning of the evolution, only the JB corresponding to the ISS steady state is populated. An adiabatic evolution will hence leave the system in the ISS of the Lindbladian at all times: $ \kket{\rho_\text{ad}(t)} = \kket{\mathcal{D}_0(t)} $ for all $ t $. In this setting, no transitions towards other Jordan blocks are allowed, since $ \bbrakket{\mathcal{E}_\alpha(t)}{\rho_\text{ad}(t)} \equiv \bbrakket{\mathcal{E}_\alpha(t)}{\mathcal{D}_0(t)} = \delta_{\alpha, 0} $. The density-matrix fidelity between the time-evolved state $ \kket{\rho(t)} $ and the ISS will therefore provide a measure of adiabaticity. It reads
\begin{equation}\label{eq:dm-fidelity}
    F(\rho_\text{ad}(s),\rho(s)) = {\left(\Tr\sqrt{\sqrt{\rho_\text{ad}(s)} \rho(s) \sqrt{\rho_\text{ad}(s)}}\right)}^2.
\end{equation}
We restrict to final times  $ \tau = \text{\SIlist{1;10}{\nano\second}} $~\footnote{We compute this adiabatic indicator in the density matrix representation.}.
	
\begin{figure}[t]
		\centering
		\includegraphics[width = \columnwidth]{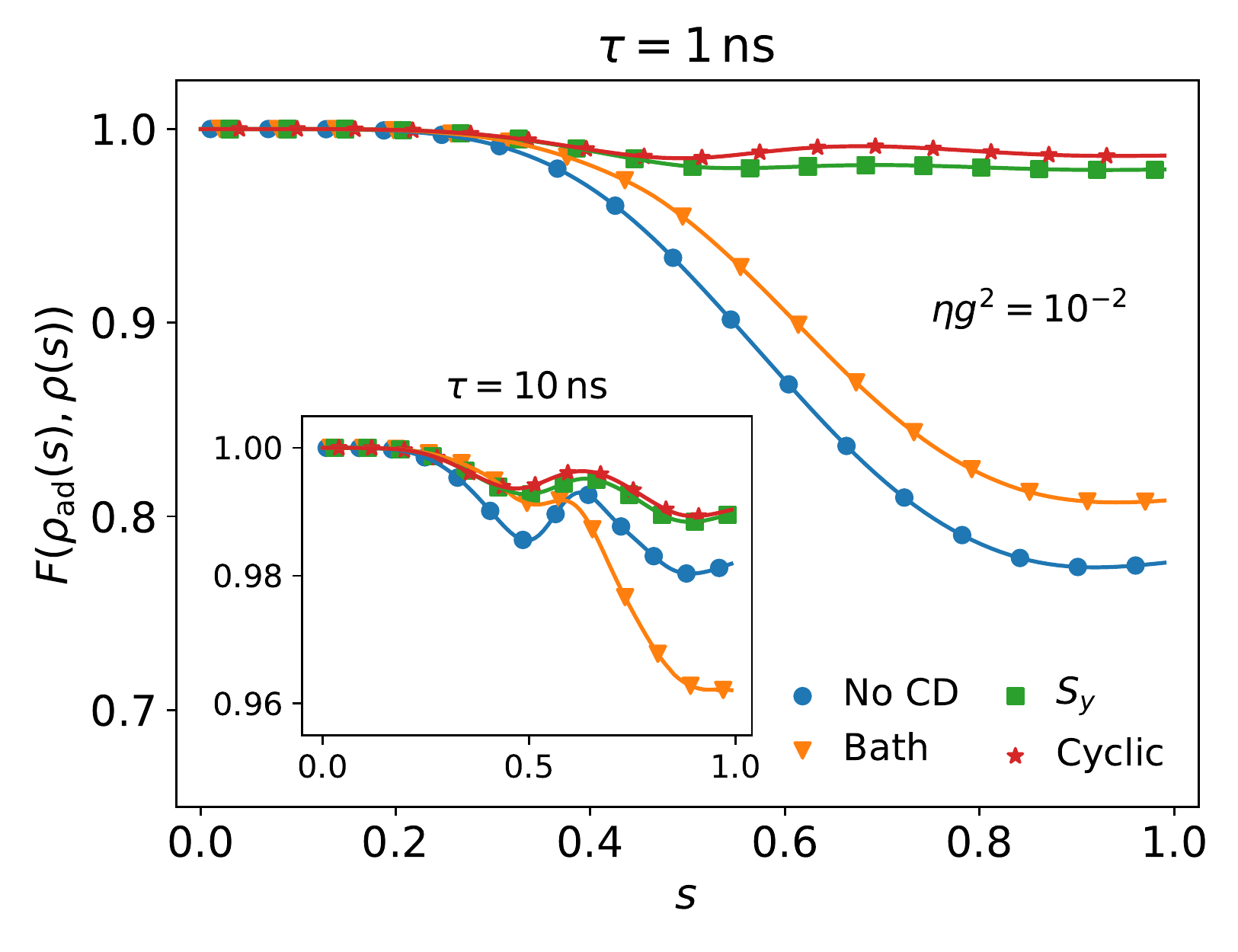}
		\caption{Fidelity between the time-evolved density matrix $ \rho(s) $ and the thermal state $ \rho_\text{ad}(s) $ as a function of $ s = t/\tau $, for $\eta g^2 = \num{1e-2}$. The different curves refer to the three ans\"atze described in the main text.}
		\label{fig:pspin-dissipative-fidelitydm-tf-1-10-ansatz-contributions-eta-0-01}
	\end{figure}	
	
In the unitary case  the $ S_y $ operator  breaks time reversal invariance and is the zeroth-order term of a number of expansions such as the local ansatz~\cite{polkovnikov:pnas}, the nested commutators ansatz~\cite{polkovnikov:nested-commutators}, or the cyclic ansatz~\cite{passarelli:cd-pspin}. When $ \tau $ is very short, the environment does not have enough time to act and the dynamics are almost unitary, thus in this regime we expect $ S_y $ to be the most relevant part of the ansatz. For longer evolutions, the environment kicks in and the dissipative part of the ansatz might play a more important role in the suppression of diabatic transitions between pairs of Jordan blocks. In the following, we will show that this is indeed not the case and a unitary ansatz for the CD superoperator is enough to decouple the system's Jordan blocks.
	
In order to highlight the different contributions to the variational CD operator, we here consider the following test Lindbladian:
	\begin{align}\label{eq:pspin-test-lindbladian-2}
		\mathbb{A}^\text{test}_s \to \mathcal{A}_s^\text{test}[\bullet] &= -i\comm{b_1 S_y + b_2 S_y^3 + b_3 (S_x S_y S_z + \text{h.\,c.})}{\bullet} \notag \\
		&\quad+ \sum_{i = 1}^{15} a_i^2 \left(\Sigma_i \bullet \Sigma_i - \frac{1}{2} \acomm{\Sigma_i \Sigma_i}{\bullet}\right).
	\end{align}
The first line of Eq.~\eqref{eq:pspin-test-lindbladian-2} describes the unitary  part and is reminiscent of the cyclic ansatz of Ref.~\cite{passarelli:cd-pspin}, which is particularly successful 
in the unitary case of the $ p $-spin model with $ p = 3 $. On the one hand, the possible experimental implementation of this 3-local term is a nontrivial task. On the other hand, terms like this are likely to appear, for instance, when using the nested commutators ansatz~\cite{polkovnikov:nested-commutators}. In addition, we only employ the cyclic ansatz as a proof of principle: as we will show later on, our results remain valid even if we consider the simpler $S_y$. 
The second line describes the (diagonal) dissipative part, including all possible basis operators for this system in the $ (4 \times 4) $-dimensional operator space barring the identity $ \Sigma_0 $, thus we have a maximum of 18 variational parameters to optimize. 

In the following calculations, we consider three special cases:  
\begin{description}
    \item[Case Bath] $ b_1 = b_2 = b_3 = 0 $ so as to consider a purely dissipative ansatz; 
    \item[Case $S_y$] $ a_i = 0 \, \forall i $ and $ b_2 = b_3 = 0 $, i.\,e., the ansatz is unitary and only includes $ S_y $;
    \item[Case Cyclic] $ a_i = 0 \, \forall i $ so as to consider the unitary cyclic ansatz.
\end{description} 
In Fig.~\ref{fig:pspin-dissipative-fidelitydm-tf-1-10-ansatz-contributions-eta-0-0001}, we plot the  fidelity between the density matrix at the time  $ s $ and the thermal density matrix at the same time, in these three cases, compared to the case with no CD superoperator, for $ \tau = \SI{1}{\nano\second} $ (main panel) and $ \tau = \SI{10}{\nano\second} $ (inset), and a coupling strength of $ \eta g^2 = \text{\numlist{1e-4}} $. The efficiency of the ansatz of Eq.~\eqref{eq:pspin-test-lindbladian-2} is mostly due to its unitary part, while the dissipative part plays a negligible role for both annealing times. As expected, for the shortest annealing time, the $ S_y $ term is responsible for the largest improvement in the fidelity. For $ \tau = \SI{10}{\nano\second} $, the fidelity is above $ 0.96 $ even in the absence of CD corrections: the instantaneous state is close to the ISS. It is remarkable that, even in this case, which should be governed by thermal processes, unitary CD superoperators allow improving the fidelity with the thermal density matrix as opposed to a purely dissipative ansatz.

On the other hand, if we increase the system-environment coupling strength to $ \eta g^2 = \num{1e-2} $, the dissipative part of the ansatz starts to play a role as shown in  Fig.~\ref{fig:pspin-dissipative-fidelitydm-tf-1-10-ansatz-contributions-eta-0-01}. 
For $ \tau = \SI{1}{\nano\second} $ (main panel), the scenario is similar to that reported in Fig.~\ref{fig:pspin-dissipative-fidelitydm-tf-1-10-ansatz-contributions-eta-0-0001} (main panel): for short annealing times, it is reasonable to expect that unitary CD superoperators would be the most effective ones since the environment will act on longer time scales. By contrast, the dissipative dynamics for $ \tau = \SI{10}{\nano\second} $ are close to being adiabatic, and the environment affects the dynamics significantly. This is evident from the inset of Fig.~\ref{fig:pspin-dissipative-fidelitydm-tf-1-10-ansatz-contributions-eta-0-01}. Here we see that the unitary part of the ansatz alone performs similarly to the case of $ \eta g^2 = \num{1e-4} $. In addition, the Cyclic and $ S_y $ ans\"atze have similar performances, as opposed to $ \eta g^2 = \num{1e-4} $. However, the striking difference is that the Bath ansatz is detrimental instead in this case: a naive minimization of the ansatz of Eq.~\eqref{eq:pspin-test-lindbladian-2} does not yield coefficients that satisfy the detailed balance Kubo-Martin-Schwinger condition~\cite{zanardi:master-equations}. Therefore, the time-evolved state departs from the ISS around the middle of the dynamics.

	\begin{figure}[t]
		\centering
		\includegraphics[width = \columnwidth]{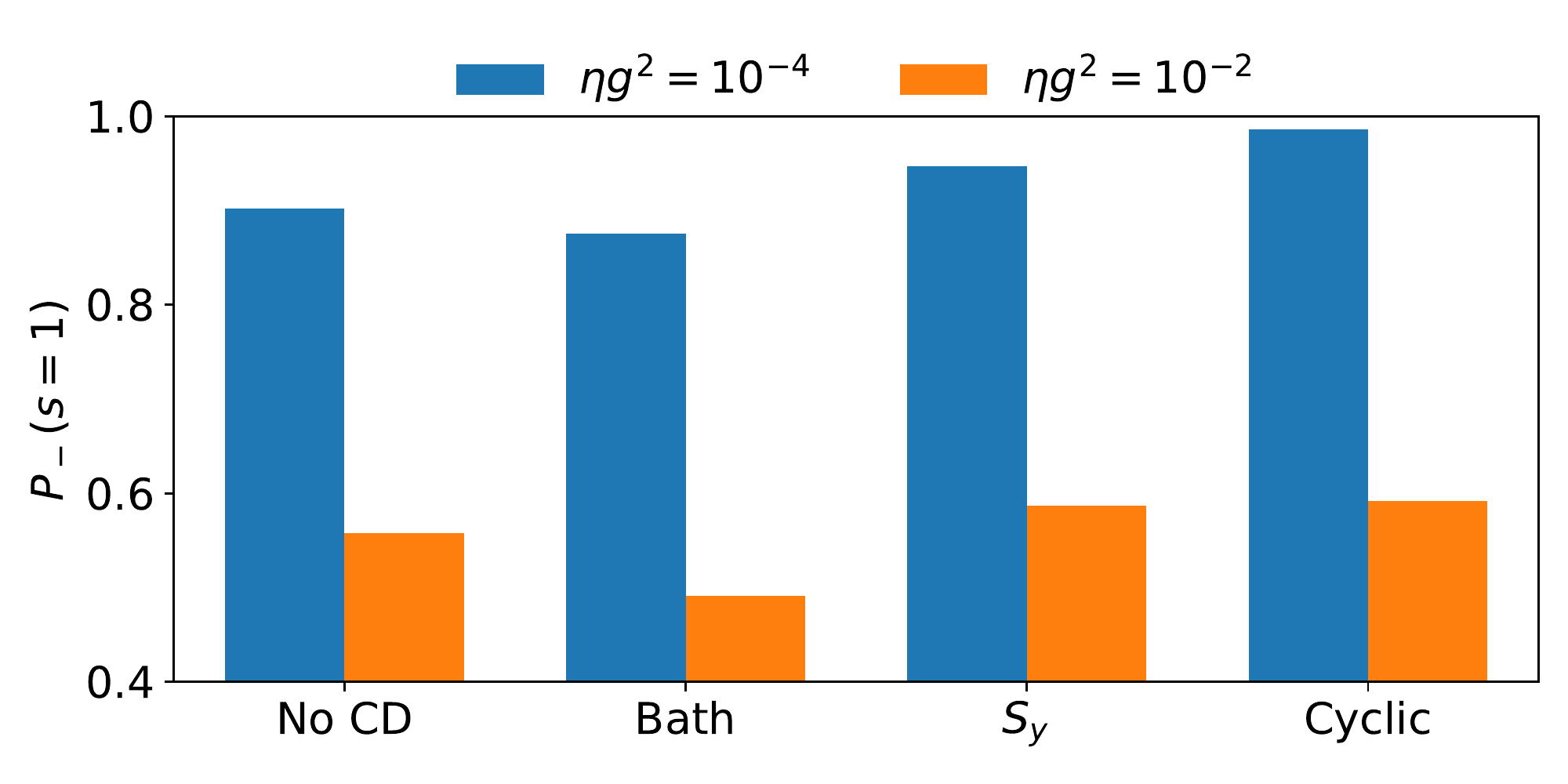}
		\caption{Ground state probability at $ t = \tau = \SI{10}{\nano\second} $ for the various ans\"atze considered in the main text, compared with the case without the CD superoperator.}
		\label{fig:pspin-dissipative-p0-tf-10-ansatz-contributions}
	\end{figure}	

Having shown that our method is able to improve open-system adiabaticity, we now prepare the $ p $-spin system in its ground state at $ t = 0 $ and evolve it using the approximate CD superoperator found previously. It can be easily done exploiting the fact that the variational minimization does not depend on the starting state.
We show that  we can additionally improve the success probability of QA, i.\,e., the ground state probability at $ t = \tau $.
Our results are summarized in Fig.~\ref{fig:pspin-dissipative-p0-tf-10-ansatz-contributions}, where we report the ground state probability at $ t = \tau = \SI{10}{\nano\second} $ for the three ans\"atze, for $ \eta g^2 = \text{\numlist{1e-4;1e-2}} $. Clearly, a unitary ansatz for the gauge potential is more geared towards the optimization of QA. The inclusion of dissipation in the ansatz negatively affects the ground state probability: a purely dissipative ansatz (Bath) decreases the success probability of QA, whereas we observe an enhancement in the GS probability for the $S_y$ and cyclic ans\"atze, consistently with known results for unitary dynamics~\cite{passarelli:cd-pspin}. Thus, we stress that no quantum channel engineering is required in this scheme: controlling the unitary evolution of a dissipative quantum system alone can improve open- and closed-system adiabaticity.

	\section{Conclusions}
	\label{sec:conclusions}
	
	In conclusion, we have formulated the search for dissipative counterdiabatic superoperators on variational grounds. We have applied our method to a relevant system for adiabatic quantum computation and we have shown that known unitary ans\"atze for the counterdiabatic gauge potential offer an excellent compromise between open- and closed-system adiabaticity in that they are able to reduce the coupling between the Jordan blocks in which Lindbladians are decomposed and, at the same time, enhance the ground state probability at the end of the dynamics.
	
Our approach can be applied if the system dynamics can be expressed in the Lindblad form, which is always the case if the Markovian approximation holds. In general, the non-Markovian limit does not admit a generic form of the dynamical equation, hence methods have to be developed case by case. This is still an open point that we leave to future analysis.
	
	\begin{acknowledgements}
	Financial support and computational resources from MUR, PON “Ricerca e Innovazione 2014-2020”, under Grant No. "PIR01\_00011 - (I.Bi.S.Co.)" are acknowledged. G.P. acknowledges support by MUR-PNIR, Grant. No. CIR01\_00011 - (I.Bi.S.Co.).
	\end{acknowledgements}
	
\appendix

\section{Unitary adiabatic theorem in the superoperator formalism}
\label{app:jordan-blocks-from-unitary}

We start from the Liouville equation for the density operator $ \rho(t) = \ket{\psi(t)}\bra{\psi(t)} $;
	\begin{equation}\label{eq:liouville}
		\dot{\rho}(t) = -\frac{i}{\hslash} \comm{H_0(t)}{\rho(t)}.
	\end{equation}
The adiabatic basis, in which $ H_0(t) $ is diagonal, allows us to expand the ket state as $ \ket{\psi(t)} = \sum_n c_n(t) \ket{d_n(t)} $. Inserting this decomposition into Eq.~\eqref{eq:liouville} and remembering that $ \dot{\rho} = \ket{\dot{\psi}}\bra{\psi} + \ket{\psi}\bra{\dot{\psi}} $, we can write
	\begin{align}\label{eq:liouville-coeffs}
		&\sum_{nm} \left[\dot{c}_n(t) c^*_m(t) + c_n(t) \dot{c}_m^*\right] \ket{d_n(t)}\bra{d_m(t)} \notag \\ &\quad+ \sum_{nm} c_n(t) c_m^*(t) \left(\ket{\dot{d}_n(t)}\bra{d_m(t)} + \text{h.\,c.}\right) \notag \\
		&\qquad = -\frac{i}{\hslash} \sum_{nm} \left[\epsilon_n(t) - \epsilon_m(t)\right] c_n(t) c_m^*(t) \ket{d_n(t)}\bra{d_m(t)}.
	\end{align}
If we now define the multi-index $ \alpha = Dn + m \in \set{0, 1, \dots, D^2-1} $, introduce the (coherence) operator basis $ \ket{d_n(t)}\bra{d_m(t)} \to \kket{\mathcal{D}_{\alpha}} $, and define the coefficient of the coherence vector as $ c_n(t) c_m^*(t) \to r_\alpha(t) $, we see that the unitary Lindbladian has a 1D Jordan representation with purely imaginary eigenvalues $ -i \left[\epsilon_n(t) - \epsilon_m(t)\right]/\hslash \to \lambda_\alpha(t) $. The Lindbladian spectrum is always degenerate here since, when $ n = m $, the corresponding $ \lambda_\alpha $ is zero.
	
Suppressing the second term on the left-hand side of Eq.~\eqref{eq:liouville-coeffs} amounts to suppressing diabatic transitions between energy eigenstates, which in turn corresponds to suppressing transitions between 1D Jordan blocks.

\section{Variational unitary CD driving as a limiting case of the 1DJF}
\label{app:cd-uni-from-open}

Here we show that the unitary variational formulation of CD driving~\cite{polkovnikov:pnas} is contained into the 1DJF open system formulation when both the Lindbladian $ \mathbb{L}_0(s) $ and the CD gauge potential $ \mathbb{A}_s^* $ are unitary maps, i.\,e.,
	\begin{gather}\label{eq:unitary-maps}
		\mathbb{L}_0(s) \to \mathcal{L}_s[\bullet] = -\frac{i}{\hslash}\comm{H_0(s)}{\bullet},\\
		\mathbb{L}_0'(s)\to \mathcal{L}'_s[\bullet] = -\frac{i}{\hslash}\comm{H_0'(s)}{\bullet},\\
		\mathbb{A}_s^* \to \mathcal{A}_s^*[\bullet] = -\frac{i}{\hslash} \comm{A_s^*}{\bullet}.
	\end{gather}
	In fact, in this case we have
	\begin{align}\label{eq:G-superoperator-unitary-maps}
		\mathbb{G}_s \to \mathcal{G}_s[\bullet] &= \mathcal{L}'_s[\bullet] - \mathcal{A}_s^*[\mathcal{L}_s[\bullet]] + \mathcal{L}_s[\mathcal{A}_s^*[\bullet]] \notag \\ &= -\frac{i}{\hslash}\comm{H_0'(s)}{\bullet} + \frac{1}{\hslash^2}\comm{A_s^*}{\comm{H_0(s)}{\bullet}} \notag \\&\quad- \frac{1}{\hslash^2} \comm{H_0(s)}{\comm{A_s^*}{\bullet}}.
	\end{align}
	Using the Jacobi identity $ [a,[b,c]] + [b, [c,a]] + [c,[a, b]] = 0 $, the last two terms yield $ [[A_s^*,H_0(s)],\bullet]/\hslash^2 $, therefore Eq.~\eqref{eq:G-superoperator-unitary-maps} is equivalent to
	\begin{align}\label{eq:G-superoperator-map}
		\mathbb{G}_s \to \mathcal{G}_s[\bullet] &= -\frac{i}{\hslash} \comm{H_0'(s) + \frac{i}{\hslash} \comm{A_s^*}{H_0(s)}}{\bullet} \notag \\ &\equiv -\frac{i}{\hslash}\comm{G_s}{\bullet},
	\end{align} 
	where $ G_s $ is defined in the main text. We now show that $ \mathbb{S} $ is minimized iff $ S $ is minimized, i.\,e., $ \mathbb{A}_s^* = \mathbb{A}_s \iff A_s^* = A_s $~\footnote{If the evolution is unitary, the resulting Lindbladian supermatrix is diagonalizable (1DJF) and the eigenvalues are $ \lambda_\alpha = -i \omega_\alpha $, where $ \omega_\alpha $ are the Bohr frequencies of the system. 
	}. To see this, it is sufficient to evaluate the matrix elements of $ \mathbb{G}_s $ in the eigenbasis of the Lindbladian, which, as shown in App.~\ref{app:jordan-blocks-from-unitary}, in case of unitary Lindbladian dynamics corresponds to the operator bases $ \bbra{\mathcal{E}_\alpha}\to\ket{d_n(t)}\bra{d_m(t)} $ with $ \alpha = Dn + m $ and $ \kket{\mathcal{D}_{\beta}}\to \ket{d_j(t)}\bra{d_k(t)} $ with $ \beta = Dj + k $. We consider the case where there are no degeneracies in the Hamiltonian spectrum so that all Bohr frequencies with different indices are nonzero. We have
	\begin{align}\label{eq:unitary-limit-variational}
		{\left(\mathbb{G}_s\right)}_{\alpha \beta} &= \mmel{\mathcal{E}_\alpha}{\mathbb{G}_s}{\mathcal{D}_\beta} = \frac{1}{D} \Tr(\mathcal{E}_\alpha^\dagger \mathcal{G}_s[\mathcal{D}_\beta]) \notag \\ &=  -\frac{i}{D\hslash} \Tr(\mathcal{E}_\alpha^\dagger \comm{G_s}{\mathcal{D}_\beta}) \notag \\ &= -\frac{i}{D\hslash} \Tr(G_s \comm{\mathcal{D}_{\beta}}{\mathcal{E}_{\alpha}^\dagger}) = \notag \\
		&= -\frac{i}{D\hslash}\sum_{ab} {\left(G_s\right)}_{ab} \comm{\mathcal{D}_{\beta}}{\mathcal{E}_{\alpha}^\dagger}_{ba} \notag \\ &= -\frac{i}{D\hslash}\sum_{ab} {\left(G_s\right)}_{ab} \left(\delta_{bj} \delta_{km} \delta_{na} - \delta_{bm} \delta_{nj} \delta_{ka}\right) = \notag \\
		&= -\frac{i}{D\hslash}\left[ {\left(G_s\right)}_{nj}\delta_{km} - {\left(G_s\right)}_{km} \delta_{nj}\right].
	\end{align}
	Four cases must be distinguished:
	\begin{enumerate}
		\item $ n \ne j $, $ k \ne m $. In this case, $ {\left(\mathbb{G}_s\right)}_{\alpha \beta} = 0 $.
		\item $ n = j $, $ k \ne m $. In this case,   $ {\left(\mathbb{G}_s\right)}_{\alpha \beta} =  \frac{i}{D\hslash} {(G_s)}_{km} $.
		\item $ n \ne j $, $ k = m $. In this case,   $ {\left(\mathbb{G}_s\right)}_{\alpha \beta} = -\frac{i}{D\hslash} {(G_s)}_{nj} $.
		\item $ n = j $, $ k = m $ (so $ \alpha = \beta $). In this case,   $ {\left(\mathbb{G}_s\right)}_{\alpha \alpha} = -\frac{i}{D\hslash} \left[{(G_s)}_{nn} - {(G_s)}_{mm}\right] $.
	\end{enumerate}
	Therefore, we have
	\begin{align}
		\mathbb{S} &= \sum_{\alpha\beta} {\lvert{\mathbb{G}_s}\rvert}^2_{\alpha\beta} = \sum_{\alpha} {\lvert\mathbb{G}_s\rvert}^2_{\alpha\alpha} + \sum_{\alpha\ne\beta} {\lvert\mathbb{G}_s\rvert}^2_{\alpha\beta}\notag \\ &= \sum_{nm} \frac{1}{D^2} {(\omega'_{nm})}^2 + \frac{2}{D\hslash^2} \sum_{m\ne k} {\lvert G_s \rvert}^2_{km}.
	\end{align}
	Therefore, $ \mathbb{S} $ is minimized when $ {(G_s)}_{km} = 0 $ for $ k \ne m $, that is, when $ A_s^* = A_s $. Conversely, if $ {(\mathbb{G}_s)}_{\alpha\beta} = 0 $ with $ \alpha \ne \beta $, then according to points 2-3 we must have $ {(G_s)}_{km} = 0 $ for $ k \ne m $, thus the unitary variational approach to CD driving of Ref.~\cite{polkovnikov:pnas} is a particular case of the more general Lindbladian formulation. 

    \section{Weak-coupling limit Lindblad equation}
	\label{app:lindblad-equation}
	
	Consider a system-bath Hamiltonian of the form $ H_{SB}(t) = H_0(t) + H_B + g U \otimes B $, where $ U $ is a system operator, $ H_B = \sum_k \omega_k b_k^\dagger b_k $ is the Hamiltonian of the bath modeled as noninteracting bosons and $ B = \sum_k (b_k + b_k^\dagger) $. The system-bath coupling strength is $g$. The weak-coupling-limit adiabatic Lindbladian of Refs.~\cite{zanardi:master-equations,albash:decoherence} reads
	\begin{align}\label{eq:ame}
		\mathcal{L}_t[\bullet] &= -i\comm{H_0(t)+H_\text{LS}(t)}{\bullet} \notag \\&+ \sum_{\omega}\gamma\bigl(\omega(t)\bigr) \left(\Gamma_\omega(t) \bullet \Gamma_\omega^\dagger(t) - \frac{1}{2} \acomm{\Gamma_\omega^\dagger(t)\Gamma_\omega(t)}{\bullet}\right),
	\end{align}
	where $ \Gamma_\omega = \sum_{a,b: \epsilon_a - \epsilon_b = \omega} \ket{\epsilon_a} \bra{\epsilon_a} U \ket{\epsilon_b} \bra{\epsilon_b} $ are Lindblad operators, the Lamb shift is $ H_\text{LS}(t) = \sum_{\omega} \zeta(\omega(t)) \Gamma_\omega^\dagger(t)\Gamma_\omega(t) $, and the (Ohmic) spectral functions read
	\begin{gather}\label{eq:spectral-functions}
		\gamma(\omega) = \frac{2\pi \eta g^2 \omega e^{-\lvert\omega\rvert/\omega_\text{c}}}{1-e^{-\beta\omega}}, \\ 
		\zeta(\omega) = \text{P.\,P.}\int_{-\infty}^{\infty
		} \frac{\gamma(\omega')}{\omega-\omega'} \, \frac{\mathrm{d}\omega'}{2\pi},
	\end{gather}
	with $ \eta g^2 $ being a dimensionless parameter related to the system-bath coupling strength ($ \eta g^2 \ll 1 $), $ \beta = 1/T $ is the inverse temperature ($ k_\text{B} = 1 $), and $ \omega_\text{c} $ is a high-frequency cutoff that we fix to $ \omega_\text{c} = \SI{8\pi}{\giga\hertz} $. Equation~\eqref{eq:ame} assumes that the Born, Markov and rotating wave approximations are valid, as a consequence of the separation between system and bath time scales~\cite{zanardi:master-equations}.

%

\end{document}